\documentclass[preprintnumbers,article,amsmath,amssymb,floatfix,10pt,onecolumn,
superscriptaddress,nofootinbib]{revtex4-2}
\usepackage{amsmath}
\usepackage{eurosym}
\usepackage{epstopdf}
\usepackage{capt-of}
\usepackage{graphicx}
\usepackage{dcolumn}

\RequirePackage{graphicx}
\RequirePackage{float}
\RequirePackage{hyperref}
\RequirePackage{amsmath}
\RequirePackage{amssymb}
\RequirePackage{mathtools}
\usepackage{color}
\usepackage{comment}
\usepackage{xcolor}

\begin{document}

\title{\textbf{Constraining anisotropic universe under $f(R,T)$ theory of gravity}}
	
\author{Lokesh Kumar Sharma}
\email{lokesh.sharma@gla.ac.in}
\affiliation{Department of Physics, GLA University, Mathura 281406, Uttar Pradesh, India}

\author{Suresh Parekh}
\email{thesureshparekh@gmail.com}
\affiliation{Department of Physics, SP Pune University, Pune 411007, Maharastra, India}

\author{Saibal Ray}
\email{saibal.ray@gla.ac.in}
\affiliation{Centre for Cosmology, Astrophysics and Space Science (CCASS), GLA University, Mathura 281406, India}

\author{Anil Kumar Yadav}
\email{abanilyadav@yahoo.co.in}
\affiliation{Department of Physics, United College of Engineering and Research, Greater Noida 201310, India.}

\begin{abstract}
\noindent \textbf{Abstract:} We try to find the possibility of a Bianchi V universe in the modified gravitational field theory of $f(R,T)$. We have considered a Lagrangian model in connection between the trace of the energy-momentum tensor $T$ and the Ricci scalar $R$. In order to solve the field equations a power law for the scaling factor was also considered. To make a comparison of the model parameters with the observational data we put constraint on the model under the datasets of the Hubble parameter, Baryon Acoustic Oscillations, Pantheon, joint datasets of Hubble parameter + Pantheon and collective datasets of the Hubble parameter + Baryon Acoustic Oscillations + Pantheon. The outcomes for the Hubble parameter in the present epoch are reasonably acceptable, especially our estimation of this $H_0$ is remarkably consistent with various recent Planck Collaboration studies that utilize the $\Lambda$-CDM model. \\

\end{abstract}

\maketitle

\section{Introduction}\label{1}
In 1998, the discovery of a late-time accelerating cosmos and the subsequent development of dark energy (DE) with repulsive pressure led to a significant advancement in cosmology~\cite{SupernovaSearchTeam:1998fmf,SupernovaSearchTeam:2004lze,SDSS:2003eyi,Calabrese:2009zza,Wang:2015tua,Zhao:2017urm}. Numerous cosmological models have been examined to comprehend the characteristics of DE~\cite{Wang:2016lxa,Peebles:2002gy,Padmanabhan:2002ji,Abdalla:2020ypg,Li:2012dt}. The cosmological constant $\Lambda$ is the most viable choice for dark energy, yet it has certain issues with its theoretical implementation~\cite{Sahni:2002kh,Garriga:2000cv,Frieman:2008sn,Nojiri:2010wj}. Our understanding of dark energy is limited to its phenomenological characteristics, which remain enigmatic and unclear: (i) DE is a cosmic fluid that violates the strong energy condition and has an equation of state parameter. (ii) DE shows a lesser clustering property than dark matter (DM) which is uniformly distributed over the universe on a large cosmological scale. Apart from the $\Lambda$-CDM model, there is also the $X$-CDM model where the dynamical dark energy is parametrized along the spatial direction and the $\phi$-CDM model where dynamical dark energy is represented by a scalar field $\phi$~\cite{Samushia:2009ib,Yashar:2008ju}. We note that the $X$-CDM model is confirmed for a considerably larger set of cosmological data, such as growth factor information, BAO distance measurements, and type Ia supernova (SN Ia) apparent magnitude observations. This encourages us to build a model of Bianchi type V space-time of the accelerating cosmos.\\

\textbf{The $f(R,T)$ theory of gravity, which was put forth by Harko et al.\cite{Harko:2011} in 2011, expresses the 
Lagrangian in terms of the trace of the stress-energy tensor $T$ and the Ricci scalar $R$. This paper introduced $f(R,T)$ gravity and explored its applications to cosmology, including explanations for cosmic acceleration and other gravitational phenomena.}  It is also evident that $f(T)$ and matter function as an effective cosmological constant. Therefore, adding a DE component to the $f(R,T)$ theory of gravity may help explain the Universe's late time acceleration. \textbf{Jamil et al. \cite{Jamil/2012} have investigated $f(R,T)$ models to explain accelerated cosmic expansion and presents various cosmological solutions. Later on, the authors of Refs. \cite{Alvarenga/2013,Sharif/2012} have described the thermodynamic implications of $f(R,T)$ gravity, focusing on cosmological thermodynamics and the validity of the generalized second law of thermodynamics. Moreover, in 2014, Harko~\cite{Harko:2014pqa} has given a generalised gravity model where the thermodynamical consequences on the matter-geometry coupling will have a definite imprint. It is worthwhile to note that Nojiri et al. \cite{Nojiri/2017} have reviewed the various modifications of gravity, including f(R,T) gravity, and examined their applications to late-time cosmic acceleration and early universe inflation.} The following works are involved with some significant uses of the $f(R,T)$ gravity theory \cite{Sharma:2018ikm,Yadav:2020vih,Sharma:2019oio,Sharma:2019hqe}. Nojiri and Odintsov \cite{Nojiri:2010wj} have studied a cosmological model where $f(R)$ is used in place of the action. Carroll et al.~\cite{Carroll:2003wy} observed that the late time acceleration of our Universe could be represented satisfactorily by cosmological models motivated by $f(R)$ gravity. The authors have looked at feasible cosmological models that meet the requirements of the Solar system test for the $f(R)$ theory of gravity~\cite{Hu:2007nk}. \textbf{Therefore, the $f(R,T)$ theory of gravity implies new interactions between geometry and matter and the door is now open to the prospect of investigating matter-geometry coupling, in which the composition of matter may have more nuanced effects on the gravitational field. In this paper, we explore an anisotropic singular universe model in $f(R,T)$ gravity with non-minimal matter geometry coupling. The field equations have been precisely solved by accounting for the scale factor power law change, which drives $\Lambda$-CDM cosmological scenario. Recently Pradhan et al. \cite{Pradhan/2023} have reconstructed a model of constant jerk parameter in $f(R, T)$ gravity which explores the late time phenomenon of the universe. Therefore, it is worthwhile to note that the $f(R,T)$ gravity theory plays an important role in evoking a comprehensive theoretical explanation of the late time accelerating universe without an assistance of unusual exotic matter or energy}.\\
 
With the help of the presence of dark energy and dark matter, the $f(R,T)$ theory of gravitation contributes significantly to the explanation of the Universe's current acceleration. Different models regarding different features of the functional form of this theory have been proposed by several scientists~\cite{Sharma:2018ikm,Alvarenga:2013syu,Sahoo:2017poz}. According to this hypothesis, the way that various matter components interact with space-time curvature is a cosmological outcome.Nevertheless, a significant contribution of EMC violation causes rapid growth in the cosmic gravity models of $f(R,T)$. The impact of EMC violation in this theory has not yet been thoroughly investigated. However, a few first investigations on celestial objects by employing this theory have been proposed in the following works~\cite{Yousaf:2016lls,Yousaf:2018jkb,Yousaf:2016awj,Yousaf:2019zcb,Das:2016mxq}. Inspired by the previous conversations, we now set out to study a cosmological model for the following specifications: (i) there will be a connection between nonminimal matter and geometry, (ii) it will be associated with the space-time of Bianchi V and (iii) here gravity will be defined as $f(R,T) = f_1(R) + f_2(R)f_3(T)$. It is noteworthy that the cosmological implications of a nonminimal coupling under the framework of $f(R,T)$ theory are explored by means of a product between $R$ and $T$ or functions of them. We shall use the simplest coupling $f_1(R) = f_2(R) = R$ and $f(T) = \zeta T$, with $\zeta$ a constant. Within the $f(R,T)$ gravity formalism, the function $f(R,T)$ has a nontrivial functional form that involves in the nonminimal matter-geometry coupling. Additionally, it benefits from the fact that $\zeta$ = 0 is the retrieval of general relativity. A few practical uses for $f(R, T) = f_1(R) + f_2(R)f_3(T)$ are available in various physical situations~\cite{Sharma:2018ikm,Yadav:2019zrw}.\\

\textbf{Among various geometrical models of the universe, Bianchi type V universe represents a natural generalization of the open universe and hence its study is important in the study of fate and dynamics of the universe with non-zero curvature \cite{Collins/1974,Coles/1994}. These models are more general and more adapted to examine certain aspects of the cosmos because it has loosen the assumption of isotropy imposed in case of FLRW model of universe. In 2004, Eriksen et al. \cite{Eriksen/2004} have investigated a cosmological phenomenon which violates the statistical isotropy of the universe due to large angle anomalies appear in CMB radiations. In the recent past, several authors have studied Bianchi-V cosmological models in different physical contexts \cite{Yadav/2011,Kumar/2011,Yadav/2012,Singh/2008}. Furthermore, the studies of the Bianchi–V cosmological models create more interest as these models contain special isotropic cases and permit arbitrarily small levels of
anisotropy at some instant of cosmic time which makes these models suitable to describes early phenomenon as well as the late time dynamics of the universe. Therefore, the homogeneous and isotropic FRW cosmological models, which are used to describe
standard cosmological models, are a particular case of Bianchi–V universe, depending on whether the negatively curved spatial geometry and possibility of forming distortion due to cosmic shear. Recently Goswami et al. \cite{Goswami/2021} have investigated a bulk viscous Bianchi type V universe and constrained its model parameters with recent H(z) and Pantheon compilation of SN Ia observational data. Some important applications of cosmological models in Bianchi V space - time are given in Refs. \cite{Yadav/2014,Quzi/2023,Gusu/2021,Barrow/2000} specially in Ref \cite{Barrow/2000}, the authors have examined the effects of anisotropic shear and implications of Bianchi type V models with cosmic microwave background (CMB) anisotropies. Moreover, Hewitt et al. \cite{Hewitt/2003} have investigated a dynamical system which approaches anisotropic Bianchi V cosmology and explored the phase space of Bianchi type V models to search its long-term behavior}. \\

The paper is structured as follows: the model and its basic formalism is described in Section \ref{2}. The method of observation analysis and observational data sets are given in Section \ref{3}. Section \ref{4} deals with the energy conditions and cosmological parameters. Finally, the findings of this paper are summarized in Section \ref{5}.

\section{Mathematical background of $f(R,T )$ gravity theory}\label{2}

The space-time for Bianchi V can be provided as
\begin{equation}
    ds^2 = -c^2dt^2 + (C^2 dz^2 + B^2 dy^2) e^{2 \alpha x} + A(t)^2dx^2, \label{eq1}  
\end{equation}
where scale factors along the $x$, $y$, and $z$ axes are denoted by $A(t)$, $B(t)$, and $C(t)$, and a constant is represented by $\alpha$. 

The action under $f(R,T)$ theory can be provided as
\begin{equation}
    S =  \frac{1}{16 \pi} \int d^4 x \sqrt{-g} f(R,T) + \int d^4 x \sqrt{-g} L_m, \label{eq2}
\end{equation}
where $g$ is the metric determinant and  $L_{m}$ is the matter Lagrangian density.

Now the given equation can be expressed as
\begin{equation}
 \begin{aligned}
   &  [f'_1 (R) + f'_2(R)f'_3(T)] R_{ij} - \frac{1}{2} f'_1(R)g_{ij} \\
   &+ (g_{ij} \nabla^i \nabla_j -  \nabla_i \nabla_j) [f'_1(R) + f'_2(R) f'_3 T] \\
     & 
    = [8 \pi + f'_2(R)f'_3 (T)] T_{ij} + f_2(R) \left[ f'_3(T)p + \frac{1}{2} f_3 (T)\right] g_{ij}, \label{eq3}
 \end{aligned}
\end{equation}
where $f(R,T)$ has been expressed earlier. Here a prime denotes derivatives with respect to the arrangement under consideration.

In connection with $f(R,T$ gravity, from Eq. (\ref{eq3}), we have 
\begin{equation}
    G_{ij} = 8 \pi \tilde{T}_{ij} = 8 \pi \left(T_{ij} + T^{\text{(ME)}}_{ij} \right), \label{eq4}
\end{equation}
where $\tilde{T}_{ij}$, $T_{ij}$ and $T^{(\text{ME})}_{ij}$ denote,   respectively, the effective energy-momentum tensor, the matter energy-momentum tensor and the additional energy term caused by the trace of the energy-momentum tensor which can be obtained as
\begin{equation}
    T^{(\text{ME})}_{ij} = \frac{\zeta R}{8 \pi} \left(T_{ij} + \frac{3\rho - 7p}{2} g_{ij} \right). \label{eq5}
\end{equation}

From Eq. (\ref{eq4}), after plugging the Bianchi identities, we get the following one
\begin{equation}
    \nabla^i T_{\text{ij}} = - \frac{\zeta R}{8 \pi} \left[\nabla^i (T_{\text{ij}} + pg_{ij}) + \frac{1}{2}g_{\text{ij}} \nabla^i (\rho - 3p)\right]. \label{eq6}
\end{equation}

For the metric (\ref{eq1}), the field equation (\ref{eq4}) can be explicitly provided as
\begin{equation}
    \frac{\ddot B}{B} +  \frac{\ddot C}{C} +  \frac{\dot B \dot C}{BC} - \frac{\alpha^2}{A^2} = -8\pi \tilde{p}, \label{eq7}
\end{equation}

\begin{equation}
    \frac{\ddot C}{C} +  \frac{\ddot A}{A} +  \frac{\dot A \dot C}{AC} - \frac{\alpha^2}{A^2} = -8\pi \tilde{p}, \label{eq8}
\end{equation}

\begin{equation}
    \frac{\ddot A}{A} +  \frac{\ddot B}{B} +  \frac{\dot A \dot B}{AB} - \frac{\alpha^2}{A^2} = -8\pi \tilde{p}, \label{eq9}
\end{equation}

\begin{equation}
    \frac{\dot A \dot B}{AB} +  \frac{\dot A \dot C }{AC} +  \frac{\dot C \dot A}{CA}  + \frac{3\alpha^2}{A^2} = 8\pi \tilde{\rho}. \label{eq10}
\end{equation}

In the above, the abbreviated symbols are as follows: $\tilde{\rho} = \rho + \rho^{(\text{ME})} = p - \frac{3 \zeta}{ 8 \pi} \left(\frac{\ddot a}{a} + \frac{\dot a^2}{a^2}\right) (3\rho - 7p)$, $\tilde{p} =  p + p^{\text{(ME)}} = p + \frac{9 \zeta}{ 8 \pi} \left(\frac{\ddot a}{a} + \frac{\dot a^2}{a^2}\right) (\rho - 3p)$ and a = $(ABC)^{\frac{1}{3}}$ which is the average scale factor.

After a simple reformulation Eqs. (\ref{eq7})--(\ref{eq10}) yield the following expression
\begin{equation}
    \frac{{(ABC)^{\ddot{}}}}{ABC} = 12\pi \left(\tilde{\rho} - \tilde{p}\right).  \label{eq11}
\end{equation}

Now, the Hubble parameter can be defined as
\begin{equation}
H = \frac{\dot a}{a}, \label{eq12}
\end{equation}
whereas the average scale factor can be provided as Sharma et al.~\cite{ Sharma:2019oio} 
\begin{equation}
a = \alpha t^{\beta} \label{eq13}
\end{equation}
where $\alpha$ and $\beta$ are two constants which should be non-zero and positive in nature. 

Again, from Eqs. (\ref{eq7}) -- (\ref{eq9}) along with Eq. (\ref{eq13}), we get 
\begin{equation}
A(t) =  \alpha t^{\beta}, \label{eq14}
\end{equation}
\begin{equation}
B(t) = \xi(\alpha t^{\beta}) \text {exp} \left[\frac{\ell t^{1-3\beta}}{\alpha^{3}(1-3\beta)}\right], \label{eq15}
\end{equation}
\begin{equation}
C(t) = \xi^{-1}(\alpha t^{\beta}) \text {exp} \left[-\frac{\ell t^{1-3\beta}}{\alpha^{3}(1-3\beta)}\right],  \label{eq16}
\end{equation}
where $\xi$ and $\ell$ incorporated above are two arbitrary constants.

\section{Observational Analysis}\label{3}

\subsection{Model parameters and observational constraints}

\noindent In order to determine the relevant model parameters, we use the following data sets to draw observational constraints. \\

\noindent \textbf{1. OHD}:
From the cosmic chronometric (CC) and BAO approach, we have collected $57~H(z)$ Observed Hubble Data (OHD) points in the interval $0\leq z\leq 2.36$. Table II contains $55$ $H(z)$ data points \cite{Lohakare/2022} while all $57 ~ H(z)$ observational data points with corresponding errors from CC (31 points), and BAO (26 points) are given in Ref. \cite{Yadav/2024JHEAP}.\\
 
\noindent\textbf{2. BAO}: 
Anisotropic Baryon Acoustic Oscillation (BAO) measurements of $D_M(z)/r_d$ and $D_H(z)/r_d$ are included in the final BAO measurements of the SDSS collaboration. Basically this span to eight different redshift intervals (in which $D_H(z)=c/H(z)$ is the Hubble distance and $D_M(z)$ is the comoving angular diameter distance). Table 3 of Ref.~\cite{eBOSS:2020yzd} compiles all the BAO-only measurements.\\

\noindent\textbf{3. Pantheon sample}: 
 The Pantheon sample was used to get distance modulus measurements of SN Ia~\cite{Scolnic/2018}. The redshift range $z \in [0.001, 2.26]$ contains 1701 light curves that correspond to 1550 distinct SN Ia events.\\ 

\noindent For the present purpose the cosmological scale factor and Hubble parameter can be expressed in the following formats:
\begin{equation}
 a = \frac{a_0}{1+z} = \alpha t^{\beta},
 \label{eq: a equation}
\end{equation}

\begin{equation}
H(z) = - \frac{1}{1+z} \frac{dz}{dt}',
\label{eq: H diff eqn}
\end{equation}
where $a_0$ is the present value of the scale factor.\\

\noindent By combining Eqs. (\ref{eq: a equation}) -- (\ref{eq: H diff eqn}) and thereafter via a few straightforward steps we can have the following Hubble parameter:
\begin{equation}
  H(z) = \beta \left(\frac{a_0}{\alpha}\right)^{-\frac{1}{\beta}} (1+z)^{\frac{1}{\beta}}.
\label{eq: model equation}
\end{equation}
From the above Eq. (\ref{eq: model equation}), the present value of the Hubble parameter (which is now obviously a constant) can be obtained as $H_0 = \beta \left(\frac{a_0}{\alpha}\right)^{-\frac{1}{\beta}}$.\\

\begin{figure}
\centering
\includegraphics[width=0.6\textwidth]{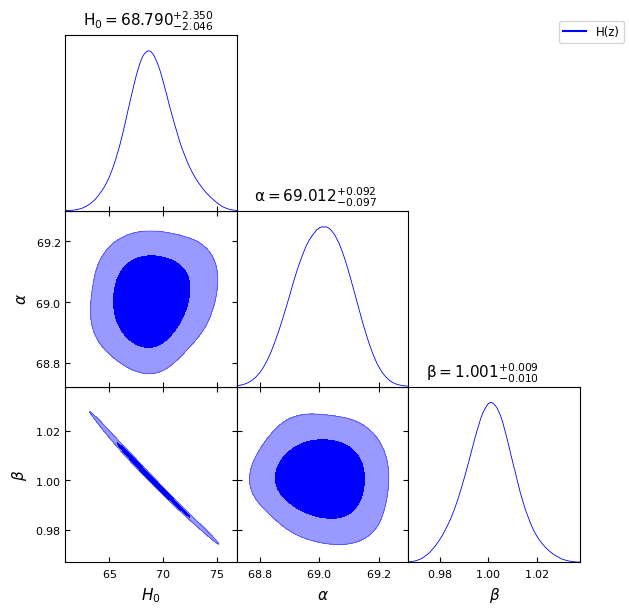}
\caption{$1D$ marginalized distribution and $2D$ contour diagrams for the present $f(R,T)$ model parameters with the $H(z)$ dataset.} \label{H(z)}
\end{figure}

\begin{figure}
\centering
\includegraphics[width=0.6\textwidth]{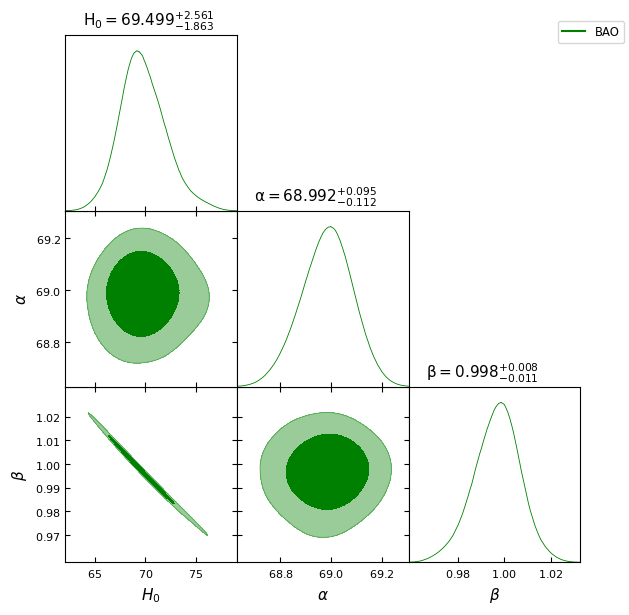}
\caption{$1D$ marginalized distribution and $2D$ contour diagrams for the present $f(R,T)$ model parameters with the BAO dataset.} \label{BAO}
\end{figure}

\begin{figure}
\centering
\includegraphics[width=0.6\textwidth]{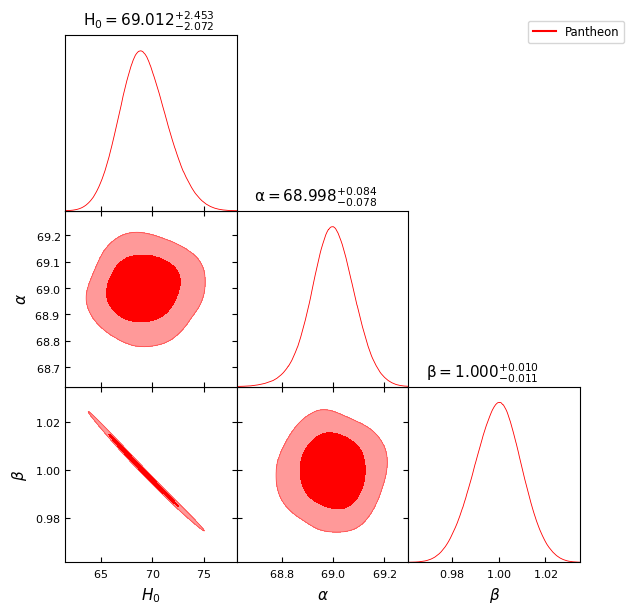}
\caption{$1D$ marginalized distribution and $2D$ contour diagrams for the present $f(R,T)$ model parameters with the Pantheon dataset.} \label{Pantheon}
\end{figure}

\begin{figure}
\centering
\includegraphics[width=0.6\textwidth]{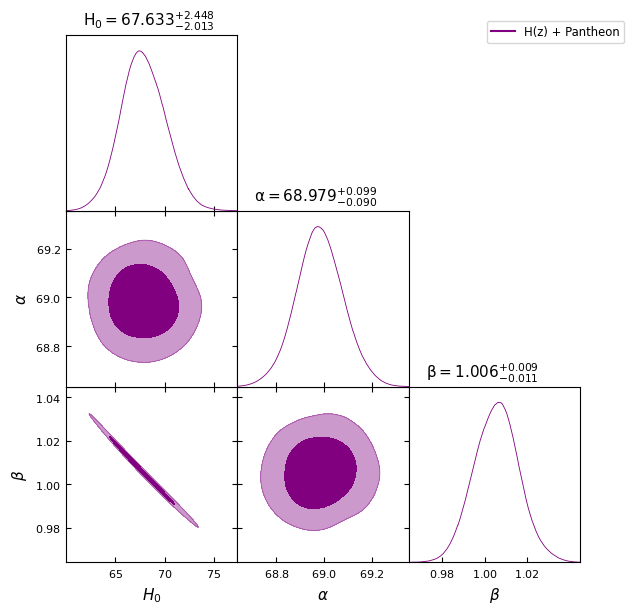}
\caption{$1D$ marginalized distribution and $2D$ contour diagrams for the present $f(R,T)$ model parameters with the combined $H(z)$ and Pantheon dataset.} \label{H(z) + Pantheon}
\end{figure}

\begin{figure}
\centering
\includegraphics[width=0.6\textwidth]{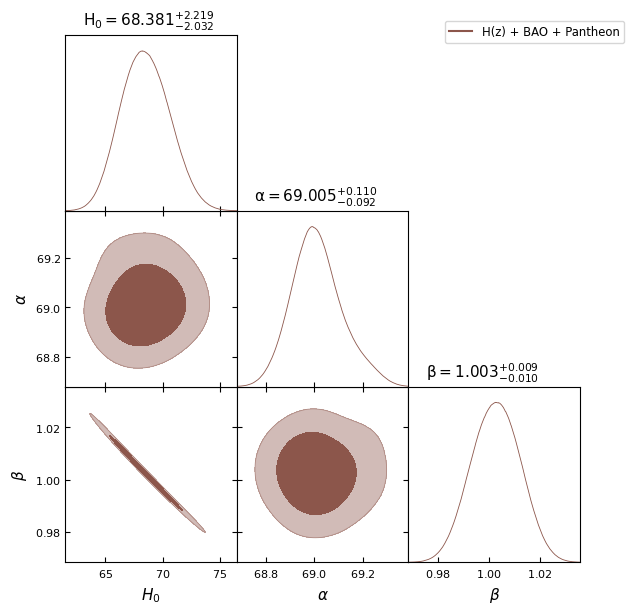}
\caption{$1D$ marginalized distribution and $2D$ contour diagrams for the present $f(R,T)$ model parameters with the combined $H(z)$, BAO and Pantheon dataset.} \label{H(z) + BAO + Pantheon}
\end{figure}

\begin{figure}
\centering
\includegraphics[width=0.6\textwidth]{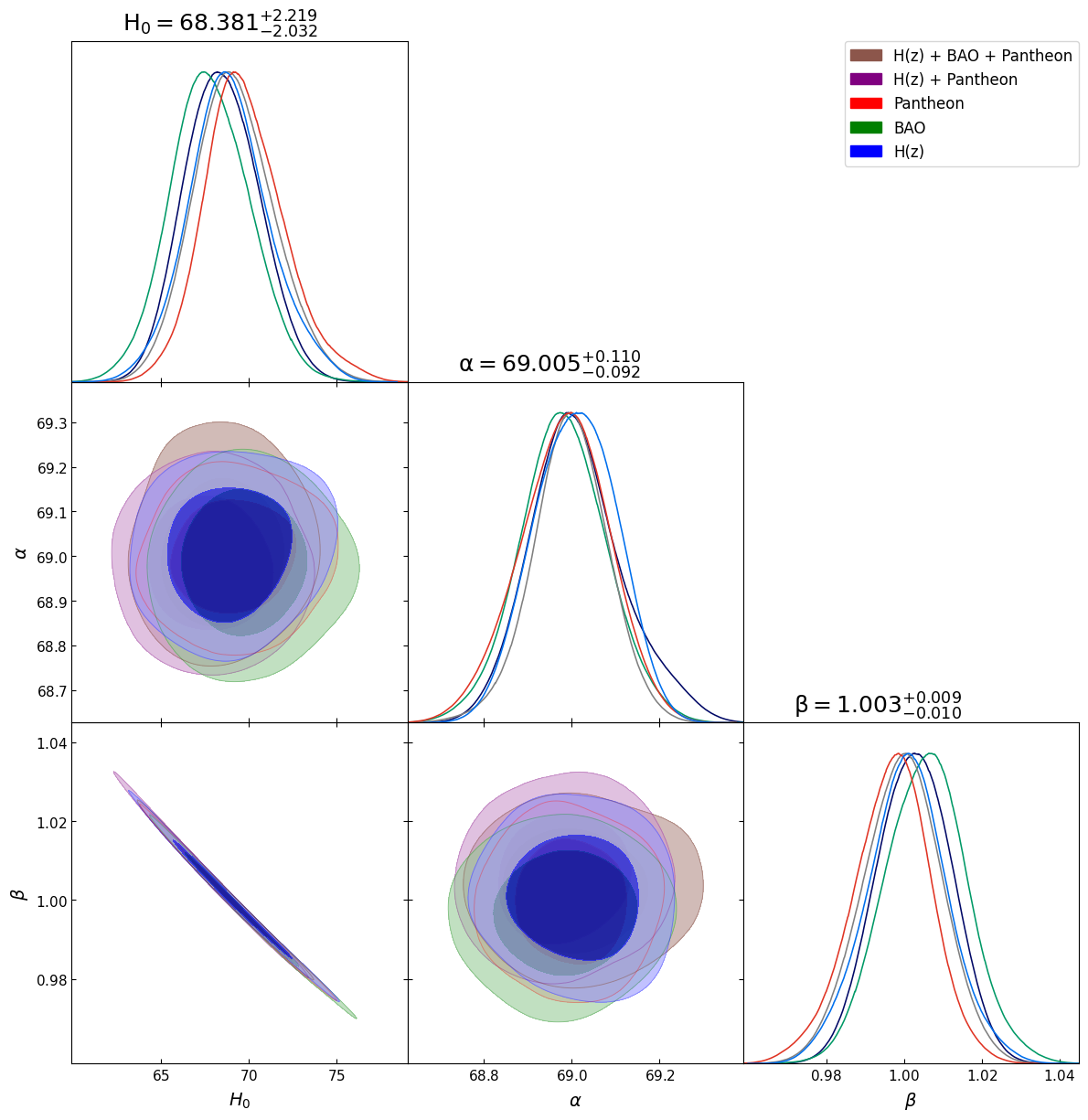}
\caption{$1D$ marginalized distribution and $2D$ contour diagrams for the present $f(R,T)$ model parameters with the combined variability across all data sets.} \label{All Combined}
\end{figure}

\begin{figure}
\centering
 \includegraphics[width=0.45\textwidth]{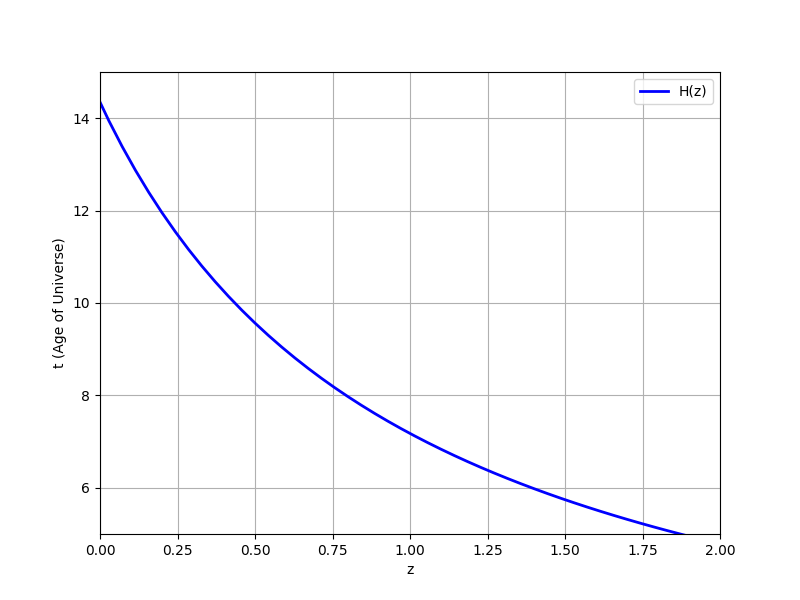}
 \includegraphics[width=0.45\textwidth]{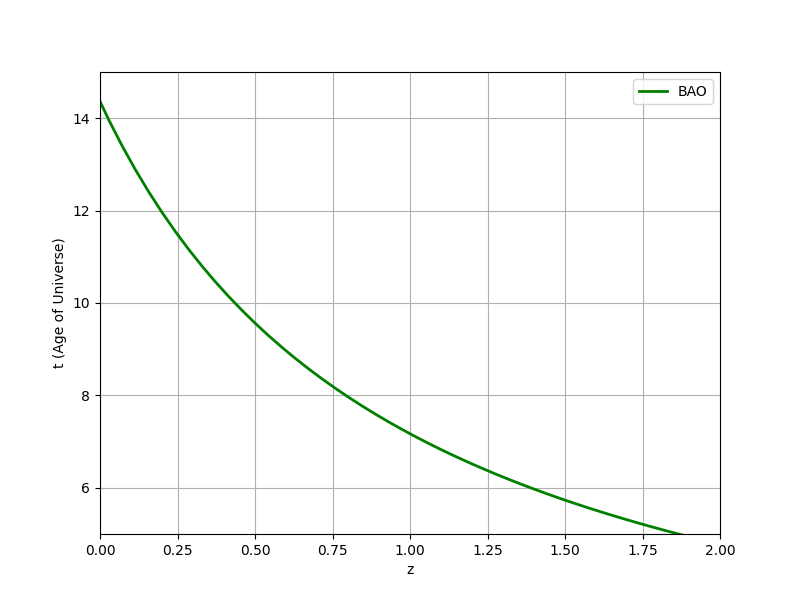}
 \includegraphics[width=0.45\textwidth]{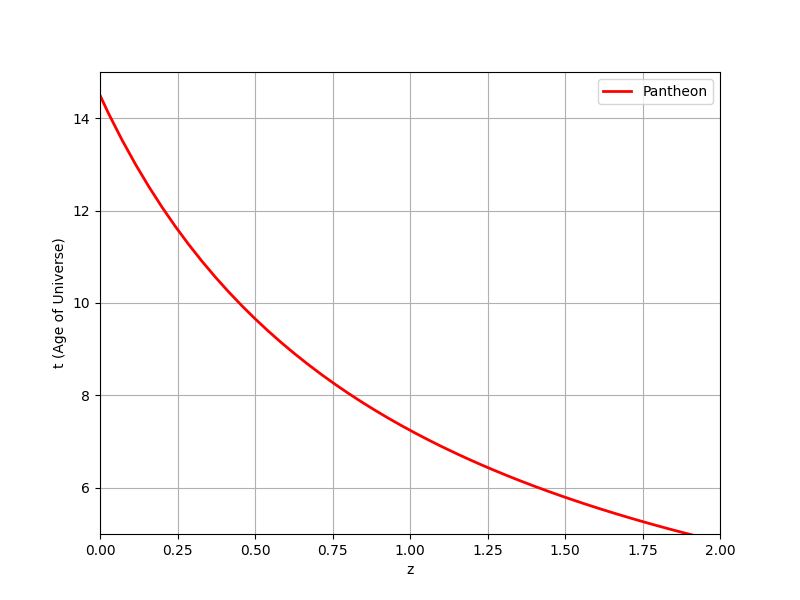}
 \includegraphics[width=0.45\textwidth]{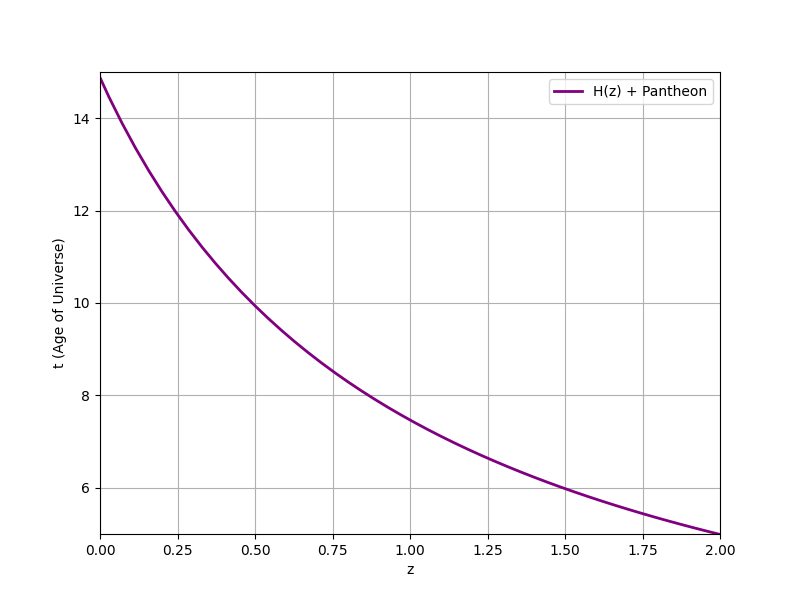}
 \includegraphics[width=0.45\textwidth]{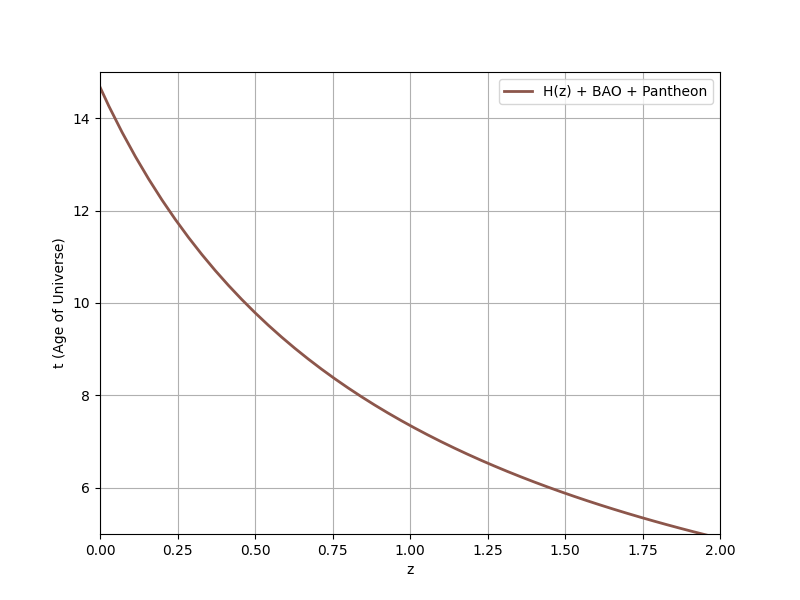}
\caption{Visual depiction showing the age of the universe as a function of redshift. Validating cosmological models and comprehending cosmic evolution depend on it. We see five graphs here for various permutations of the  data sets among $H(z)$, BAO and Pantheon.}
  \label{fig 7}
\end{figure}

\noindent Figures \ref{H(z)}$-$ \ref{All Combined} depict $1D$ marginalized distribution and $2D$ contour diagrams for the derived model and the estimated value of the parameters extracted from OHD, BAO, Pantheon compilation of SN Ia and joined data sets respectively. The parametric values (viz. $H_0$, $\alpha$, $\beta$ and the age of Universe in Gyr $t$) obtained from different data sets due to MCMC and Bayesian analysis are tabulated in Table \ref{tab:57datapt}. The graphical representation of cosmic age over redshift in exhibited in Fig.  \ref{fig 7}. Five plots are for different combinations of $H(z)$, BAO, Pantheon data sets and their combinations. The Age of universe at $z = 0$ is mentioned in Table \ref{tab:57datapt}. 

\begin{table*}
\caption{Parametric values (viz. $H_0$, $\alpha$, $\beta$ and the age of Universe (Gyr) $t$) obtained from different data sets due to MCMC and Bayesian analysis where $H_1 =H(z) + Pantheon$ and $H_2 = H(z) + Pantheon + BAO$.}
\centering
\label{tab:57datapt}
\renewcommand{\arraystretch}{1.0}
\begin{tabular}{lccccc}
\hline\\
\textbf{Parameter} & \textbf{H(z)} & \textbf{BAO} & \textbf{Pantheon} & \textbf{$H_1$} & \textbf{$H_2$} \\\\
\hline
$H_0$  & $68.790^{+2.350}_{-2.046}$ & $69.499^{+2.561}_{-1.863}$ & $69.012^{+2.453}_{-2.072}$ & $67.633^{+2.448}_{-2.013}$ & $68.381^{+2.219}_{-2.032}$\\\hline
$\alpha$ & $69.012^{+0.092}_{-0.097}$ & $68.992^{+0.095}_{-0.112}$ & $68.998^{+0.084}_{-0.078}$ & $68.979^{+0.099}_{-0.090}$ & $69.005^{+0.110}_{-0.092}$ \\\hline
$\beta$ & $1.001^{+0.009}_{-0.010}$ & $0.998^{+0.008}_{-0.011}$ & $1.000^{+0.010}_{-0.011}$ & $1.006^{+0.009}_{-0.011}$ & $1.003^{+0.009}_{-0.010}$ \\\hline
$t$ & $14.3430291$ & $14.3599188$ & $14.4902335$ & $14.8743956$ & $14.6678171$ \\
\hline
\end{tabular}
\end{table*}

\section{Energy conditions and cosmological Parameters}\label{4}

\subsection{Energy Conditions}

\noindent The energy conditions, analogously establish cosmic laws that elucidate the distribution of the matter and energy throughout the cosmos which can be derived from Einstein's gravitational equations. Thus we obtain certain conditions, viz. Weak Energy Condition (WEC), Null Energy Condition (NEC), Strong Energy Condition (SEC) and Dominant Energy Condition (DEC) which reveal the distribution of matter and energy in the cosmic space.

\begin{enumerate}
    \item WEC: $ \rho \geq 0 $\\
        According to the WEC, energy can nowhere be negative or zero. This requirement keeps the laws of the cosmos impartial and constant. Using the parameters discovered using the Bayesian Analysis, Fig. \ref{fig:WEC} depicts the WEC's nature for our model.

    \item NEC: $\rho - p \geq 0 $ \\
        The NEC deals with light. It asserts that energy is a necessary component of all space transit for light and cannot exist in a vacuum. This constraint prevents odd occurrences from occurring and maintains the physics of the cosmos. Figure \ref{fig:NEC} illustrates the characteristics of NEC for our model utilizing the Bayesian Analysis's parameters.

    \item SEC: $ \rho + 3p \geq 0 $\\
        Similar to a more rigid counterpart of the NEC, the SEC which guarantees that the energy cannot be negative and sets a boundary for how objects respond to gravity. The cosmos benefits from having this rule in place. Figure \ref{fig:SEC} depicts the characteristics of SEC for our model utilizing the Bayesian analysis.

    \item DEC: $\rho + p \geq 0 $\\
        The DEC guarantees that the energy is not just non-negative but also that its distribution cannot be too volatile. It is comparable to asserting that the energy cannot go out of control or travel faster than light. Based on the Bayesian Analysis DEC is shown in Fig. \ref{fig:DEC}.
\end{enumerate}

\begin{figure}
  \centering
  \includegraphics[width=0.5\textwidth]{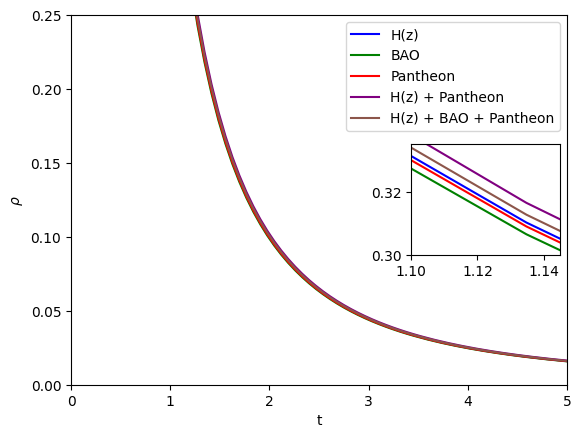}
  \caption{Illustration of the dynamic variation of the energy density over time under various parametric conditions derived from the combined H(z), BAO and Pantheon data sets.}
  \label{fig:rho}
\end{figure}

\begin{figure}
  \centering
  \includegraphics[width=0.5\textwidth]{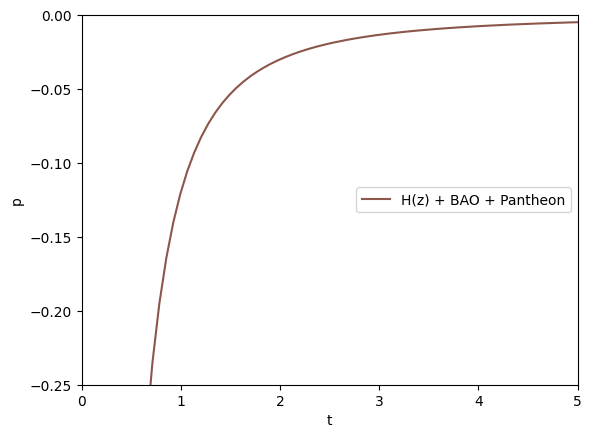}
  \caption{Using the combined H(z), BAO, and Pantheon data sets, this figure illustrates the dynamic pressure variation over time under different parameter circumstances.}
  \label{fig:p}
\end{figure}

\noindent All the above-mentioned energy conditions combinedly are as shown in the Fig. \ref{fig:energy}. Figure \ref{fig:rho} displays the distribution of energy density $\rho$ versus time $t$, while Fig. \ref{fig:p} displays the distribution of pressure. Our results demonstrate that, except for SEC the others, e.g. NEC, WEC and DEC are fulfilled. The late time rapid expansion of the universe justifies violation of SEC. The $f(R,T)$ theory of gravity having contribution from the trace energy $T$ can suitably explain the late time accelerated phase of the universe without invoking any need for the cosmological constant or exotic dark energy as the background agent for the phenomenon.

\begin{figure}
  \centering
  \includegraphics[width=0.5\textwidth]{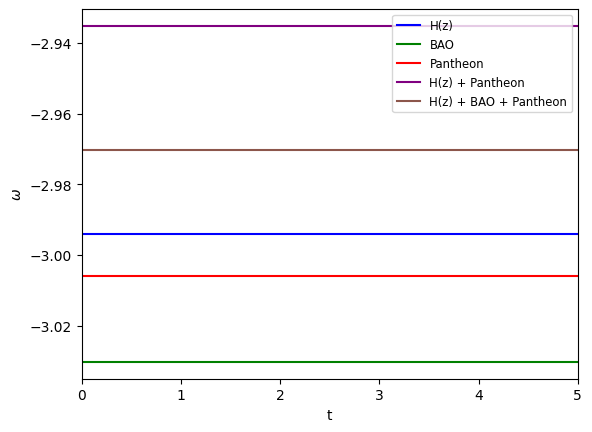}
  \caption{Graphical presentation of equation of state parameter vs time.}
  \label{fig:sep_energy}
\end{figure}

\begin{figure}
  \centering
  \includegraphics[width=0.5\textwidth]{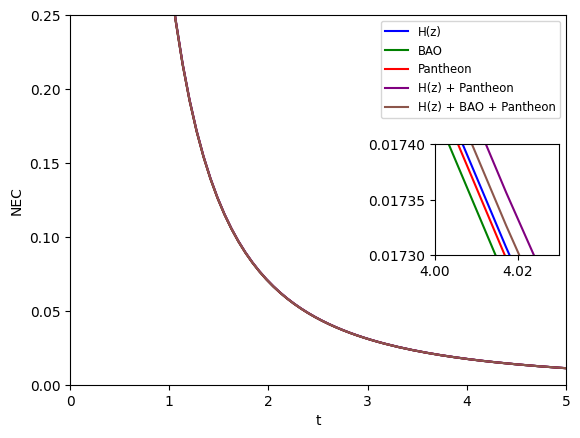}
  \caption{Graphical presentation of NEC vs time for all dataset combinations.}
  \label{fig:NEC}
\end{figure}

\begin{figure}[htbp]
  \centering
  \includegraphics[width=0.5\textwidth]{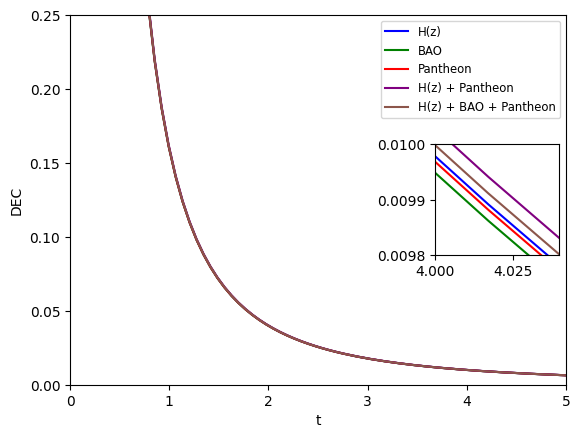}
  \caption{Graphical presentation of DEC vs time for all dataset combinations.}
  \label{fig:DEC}
\end{figure}

\begin{figure}[htbp]
  \centering
  \includegraphics[width=0.5\textwidth]{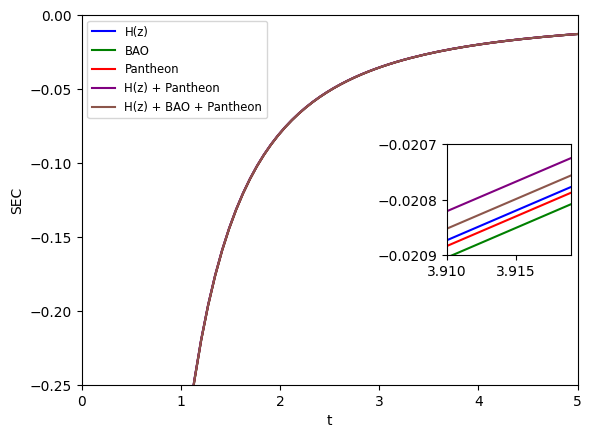}
  \caption{Graphical presentation of SEC vs time for all dataset combinations.}
  \label{fig:SEC}
\end{figure}

\begin{figure}[htbp]
  \centering
  \includegraphics[width=0.5\textwidth]{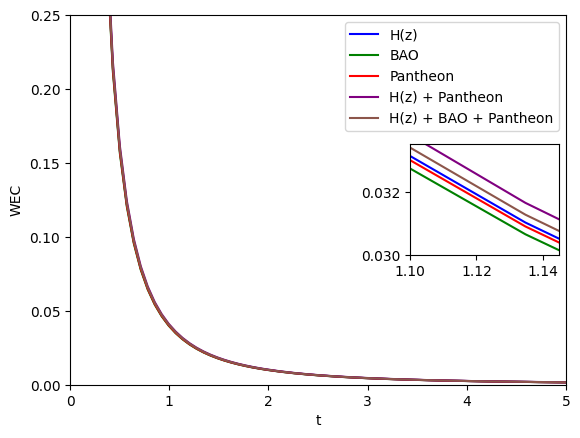}
  \caption{Graphical presentation of WEC vs time for all dataset combinations.}
  \label{fig:WEC}
\end{figure}
\begin{figure}
  \centering
  \includegraphics[width=0.5\textwidth]{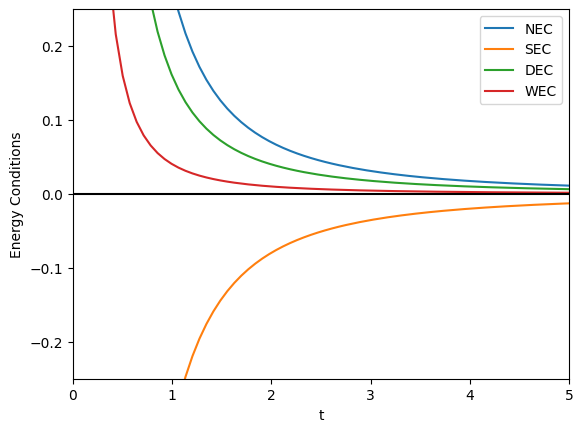}
  \caption{Graphical presentation of all the energy conditions vs time.}
  \label{fig:energy}
\end{figure}


\subsection{State Finder Diagnotics}

\noindent The state finder diagnostics are a kind of pathological testings which can assist to find out  the puzzles of dark energy and thus the history of the cosmos. These diagnostics are based on the $r$ and $s$ parameters and we can comprehend the evolution of the cosmos by using these factors in a better way as that provide information on the expansion of the universe and the elements that make it up. These dimensionless parameters help us understand the fundamental dynamics of the cosmos having definitions as follows:
\begin{equation}
    r = \frac{\ddot{\dot{a}}}{aH^3},
\end{equation}

\begin{equation}
    s = \frac{(-1+r)}{3(-\frac{1}{2} + q)}.
\end{equation}

The above two equations for the model under consideration take the following forms:
\begin{equation}
r = q(1+2q),
\end{equation}

\begin{equation}
s = \frac{2}{3}(q+1).
\end{equation}

\begin{figure}
  \centering
   \includegraphics[width=0.30\textwidth]{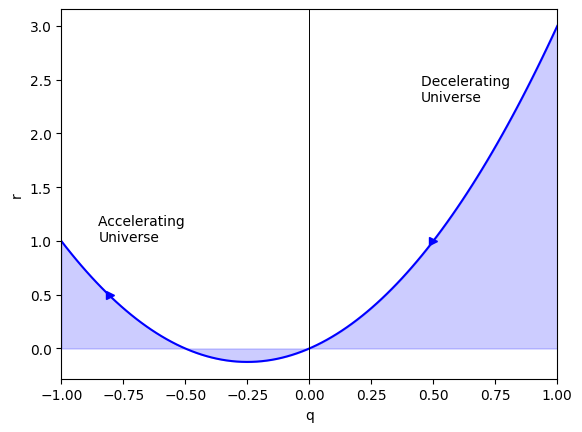}
   \includegraphics[width=0.30\textwidth]{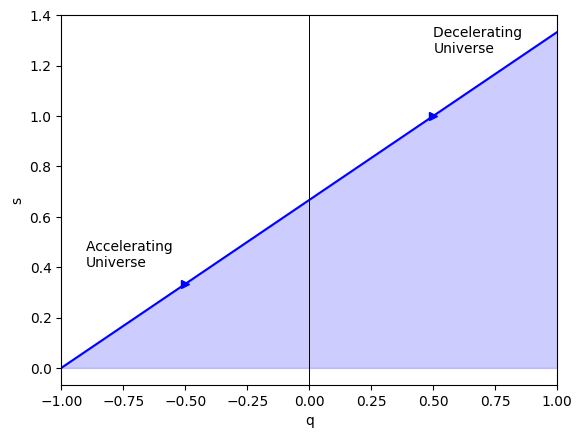}
   \includegraphics[width=0.30\textwidth]{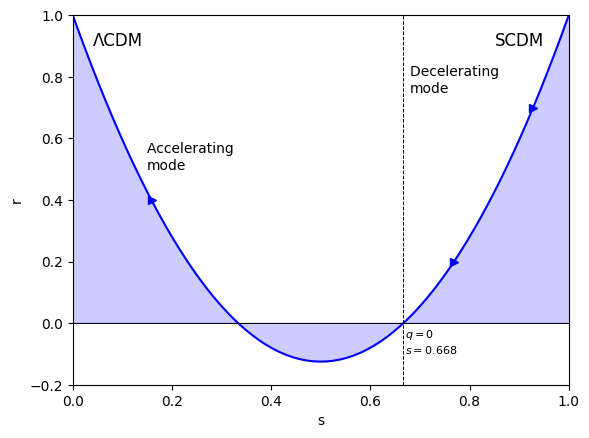}
  \caption{Graphical presentations of the state finder parameters $r-q$, $s-q$ and $r-q$.}
  \label{fig:statfind}
\end{figure}

The scale factor trajectories in the resulting model may be shown in Fig. \ref{fig:statfind} to follow a specific set of paths. Interestingly, our strategy is consistent with the results for the cosmic diagnostic pair from power law cosmology.

\subsection{Om(z) parameter}

\noindent The state finder parameters $r - s$ and the $Om(z)$ diagnostic are usually applied to analyse various dark energy ideas. Basically, the $Om(z)$ parameter is deployed when the Hubble parameter and the cosmic redshift merge in a suitable way. The $Om(z)$ parameter in the alternative gravity is read as
\begin{equation}
Om(z) = \frac{[\frac{H(z)}{H_0}]^2 - 1}{(1+z)^3 - 1}.
\end{equation}

\noindent The Hubble parameter is shown by $H_0$ in this case. Shahalam et al.~\cite{Shahalam2015} claim that the negative, zero and positive values of $Om(z)$, respectively, represent the quintessence ($\omega \ge -1$), $\Lambda$-CDM and phantom ($\omega \le-1$) dark energy theories. However, for the present model, we obtain this parameter as follows:
\begin{equation}
    Om(z) = \frac{(1+z)^{2/b}-1}{(1+z)^3-1}.
\end{equation}

\begin{figure}
  \centering
  \includegraphics[width=0.6\textwidth]{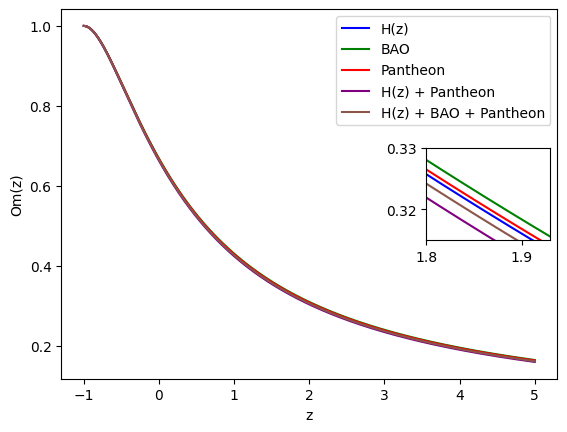}
  \caption{Graphical presentations of $Om(z)$ with $z$ for combined dataset based on the $\beta$ values obtained from each dataset.}
  \label{fig:om(z)}
\end{figure}

\subsection{Jerk Parameter}

\begin{figure}
  \centering
  \includegraphics[width=0.6\textwidth]{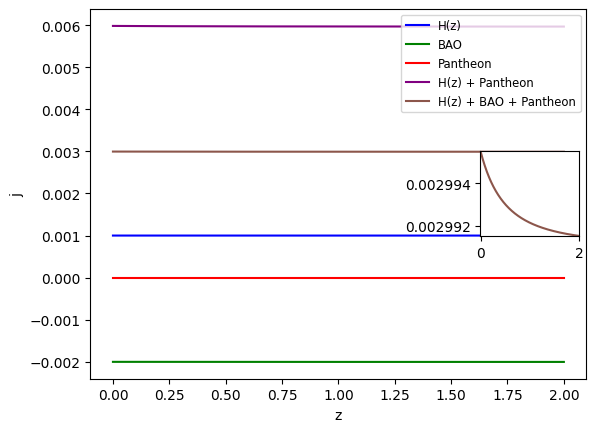}
  \caption{Graphical presentation of the jerk parameter vs redshift. Significant values of $J$ at $z = 0$ are as follows: (i) for H(z) = 0.0009995 $s^{-3}$, (ii) for BAO = -0.002002 $s^{-3}$, (iii) for Pantheon = 0.0 $s^{-3}$, (iv) for $H(z)$ + Pantheon = 0.0059817 $s^{-3}$ and (v) for $H(z)$ + BAO + Pantheon = 0.0029954 $s^{-3}$}.
  \label{fig:jerk}
\end{figure}

\noindent After the Hubble and deceleration parameters, the jerk parameter is one which can provide deeper insights into the universe's history. The universe's acceleration is tracked throughout time by the jerk parameter, indicated by the symbol $j$. The acceleration is accelerating and speeding up phase when $j$ is positive ($j \ge 0$) whereas the acceleration is said to slow down if $j$ is negative ($j \le 0$). However, different dark energy theories forecast various jerk parameter values and thus scientists can learn more about dark energy  by measuring $j$ and comparing it to these predictions. The jerk parameter contributes to the improvement of our models of the cosmos when paired with other cosmic factors. It puts these models to a rigorous examination to ensure they truly reflect what we observe in the physical cosmos. In this context we note that Fig. \ref{fig:jerk} depicts the fluctuation of $j$ with respect to redshift $z$ for our model and the derived parameters.

\section{Conclusion}\label{5}

\noindent In the present work our aim was to find the possibility of a Bianchi V type Universe under $f(R,T)$ gravity theory. For this purpose we have proposed for a viable Lagrangian model based on the connection between the trace of the energy-momentum tensor $T$ and the Ricci scalar $R$. However, in order to solve the related field equations we have considered a power law for the scaling factor.\\

\noindent The investigation provides the following salient features:\\

\noindent (i) To compare the model parameters with the observational data we impose constraint on the model using the data sets from the Hubble parameter $H(z)$, BAO, Pantheon, combined $H(z)$ + Pantheon, and combined $H(z)$ + BAO + Pantheon. The outcomes for $H_0$ under our $f(R,T)$ gravity model are as follows: $H_0 = 68.790^{+2.250}_{-2.046}$ km s$^{-1}$ Mpc$^{-1}$, $H_0 = 69.499^{+2.561}_{-1.863}$ km s$^{-1}$ Mpc$^{-1}$, $H_0 = 69.102^{+2.453}_{-2.072}$ km s$^{-1}$ Mpc$^{-1}$, $H_0 = 67.633^{+2.448}_{-2.013}$ km s$^{-1}$ Mpc$^{-1}$, $H_0 = 68.381^{+2.219}_{-2.032}$ km s$^{-1}$ Mpc$^{-1}$, respectively. \\

\noindent (ii) Our estimation of $H_0$ is remarkably consistent with various recent Planck Collaboration studies that utilize the $\Lambda$-CDM model. \\

\noindent (iii) We have addressed the energy conditions, $Om(z)$ analysis and cosmographical parameters which are equivocally very responsive and informative for physical viability of our model. \\

\noindent (iv) Graphical presentations of different model parameters are shown in several figures which are very informative in their nature. Among all these the Fig. \ref{fig:sep_energy}, considering the relationship between the state parameter of the equation and the passage of time implies that dark energy plays a role in the accelerated expansion of the universe, while its exact nature varies over time. This finding could have intriguing cosmological implications.\\

\noindent Overall statistical analysis and conclusion can be put forward as follows: the obtained results demonstrate that the proposed model mostly agrees with the observational signatures of the cosmological scenario within a certain range of restrictions. \\

\section*{Data Availability}             
\noindent This research did not yield any new data.. 

\section*{Conflict of Interest} 
\noindent The authors declare no conflict of interest.

\section*{Acknowledgement}
\noindent We are very much grateful to the honorable referee and to the editor for the illuminating
suggestions that have significantly improved our work in terms of research quality and presentation.
SR would like to express his gratitude to the ICARD facilities at CCASS, GLA University, Mathura, as well as the Visiting Research Associateship Programme at the Inter-University Centre for Astronomy and Astrophysics (IUCAA) in Pune, India. AKY is grateful to Dr. Nafis Ahmad, King Khalid University for fruitful discussions. The author (AKY) extends his appreciation to the Deanship of Scientific Research at King Khalid University, Saudi Arabia for funding this work through large group Research Project under grant number RGP. 2/315/45.\\


\begin{thebibliography}{100}

\bibitem{SupernovaSearchTeam:1998fmf} A.~G.~Riess \textit{et al.} [Supernova Search Team], Observational evidence from supernovae for an accelerating universe and a cosmological constant, Astron. J. \textbf{116}, 1009-1038 (1998)

\bibitem{SupernovaSearchTeam:2004lze} A.~G.~Riess \textit{et al.} [Supernova Search Team], Type Ia supernova discoveries at z \ensuremath{>} 1 from the Hubble Space Telescope: Evidence for past deceleration and constraints on dark energy evolution, Astrophys. J. \textbf{607}, 665-687 (2004)

\bibitem{SDSS:2003eyi} M.~Tegmark \textit{et al.} [SDSS], Cosmological parameters from SDSS and WMAP, Phys. Rev. D \textbf{69}, 103501 (2004)

\bibitem{Calabrese:2009zza} E.~Calabrese, M.~Migliaccio, L.~Pagano, G.~De Troia, A.~Melchiorri and P.~Natoli, Cosmological constraints on the matter equation of state, Phys. Rev. D \textbf{80}, 063539 (2009)

\bibitem{Wang:2015tua} Y.~Wang and M.~Dai, Exploring uncertainties in dark energy constraints using current observational data with Planck 2015 distance priors, Phys. Rev. D \textbf{94}, no.8, 083521 (2016)

\bibitem{Zhao:2017urm} M.~M.~Zhao, D.~Z.~He, J.~F.~Zhang and X.~Zhang, Search for sterile neutrinos in holographic dark energy cosmology: Reconciling Planck observation with the local measurement of the Hubble constant, Phys. Rev. D \textbf{96}, no.4, 043520 (2017)

\bibitem{Wang:2016lxa} B.~Wang, E.~Abdalla, F.~Atrio-Barandela and D.~Pavon, Dark Matter and Dark Energy Interactions: Theoretical Challenges, Cosmological Implications and Observational Signatures, Rept. Prog. Phys. \textbf{79}, 096901 (2016)

\bibitem{Peebles:2002gy} P.~J.~E.~Peebles and B.~Ratra, The Cosmological Constant and Dark Energy, Rev. Mod. Phys. \textbf{75}, 559-606 (2003)

\bibitem{Padmanabhan:2002ji} T.~Padmanabhan, Cosmological constant: The Weight of the vacuum, Phys. Rept. \textbf{380}, 235-320 (2003)

\bibitem{Abdalla:2020ypg} E.~Abdalla and A.~Marins, The Dark Sector Cosmology, Int. J. Mod. Phys. D \textbf{29}, 2030014 (2020)

\bibitem{Li:2012dt} M.~Li, X.~D.~Li, S.~Wang and Y.~Wang, Dark Energy: A Brief Review, Front. Phys. (Beijing) \textbf{8}, 828-846 (2013)

\bibitem{Sahni:2002kh} V.~Sahni, The Cosmological constant problem and quintessence, Class. Quant. Grav. \textbf{19}, 3435-3448 (2002)

\bibitem{Garriga:2000cv} J.~Garriga and A.~Vilenkin, Solutions to the cosmological constant problems, Phys. Rev. D \textbf{64}, 023517 (2001)

\bibitem{Frieman:2008sn} J.~Frieman, M.~Turner and D.~Huterer, Dark Energy and the Accelerating Universe, Ann. Rev. Astron. Astrophys. \textbf{46}, 385-432 (2008)

\bibitem{Nojiri:2010wj} S.~Nojiri and S.~D.~Odintsov, Unified cosmic history in modified gravity: from $F(R)$ theory to Lorentz non-invariant models, Phys. Rept. \textbf{505}, 59-144 (2011)

\bibitem{Samushia:2009ib} L.~Samushia and B.~Ratra, Constraining dark energy with gamma-ray bursts, Astrophys. J. \textbf{714}, 1347-1354 (2010)

\bibitem{Yashar:2008ju} M.~Yashar, B.~Bozek, A.~Abrahamse, A.~Albrecht and M.~Barnard, Exploring Parameter Constraints on Quintessential Dark Energy: the Inverse Power Law Model, Phys. Rev. D \textbf{79}, 103004 (2009)

\bibitem{Harko:2011}Harko, Tiberiu, Francisco SN Lobo, Shin’ichi Nojiri, and Sergei D. Odintsov, $f(R,T)$ gravity, Phys. Rev. D \textbf{84}, 024020 (2011)

\bibitem{Jamil/2012} M. Jamil, D. Momeni, M. Raza, and R. Myrzakulov, Reconstruction of some cosmological models in $f(R,T)$ gravity, European Physical Journal C \textbf{72}, 1999 (2012) 

\bibitem{Alvarenga/2013} F. G. Alvarenga, M. J. S. Houndjo, A. V. Monwanou and J. B. C. Orou, Thermodynamics in $f(R,T)$ gravity, Journal of Modern Physics, \textbf{4}, 130 (2013) 

\bibitem{Sharif/2012} M. Sharif and M. Zubair, Thermodynamics in $f(R,T)$ Theory of Gravity, Journal of the Physical Society of Japan, \textbf{81}, 114005 (2012)

\bibitem{Harko:2014pqa} T.~Harko, Thermodynamic interpretation of the generalized gravity models with geometry - matter coupling, Phys. Rev. D \textbf{90}, no.4, 044067 (2014)

\bibitem{Nojiri/2017} S. Nojiri, S. D. Odintsov and V. K. Oikonomou, Modified Gravity Theories on a Nutshell: Inflation, Bounce and Late-time Evolution, Physics Reports \textbf{692}, 1-104 (2017)

\bibitem{Sharma:2018ikm} L.~K.~Sharma, A.~K.~Yadav, P.~K.~Sahoo and B.~K.~Singh, Non-minimal matter-geometry coupling in Bianchi I space-time, Res. Phys. \textbf{10}, 738-742 (2018)

\bibitem{Yadav:2020vih} A.~K.~Yadav, L.~K.~Sharma, B.~K.~Singh and P.~K.~Sahoo, Existence of bulk viscous universe in $f(R,T)$ gravity and confrontation with observational data, New Astron. \textbf{78}, 101382 (2020)

\bibitem{Sharma:2019oio} L.~K.~Sharma, A.~K.~Yadav and B.~K.~Singh, Power-law solution for homogeneous and isotropic universe in $f(R,T)$ gravity, New Astron. \textbf{79}, 101396 (2020)

\bibitem{Sharma:2019hqe} L.~K.~Sharma, B.~K.~Singh and A.~K.~Yadav, Viability of Bianchi type V universe in $f(R,T)= f_{1}(R)+f_{2}(R)f_{3}(T)$  gravity, Int. J. Geom. Meth. Mod. Phys. \textbf{17}, 2050111 (2020)

\bibitem{Nojiri:2010wj} S.~Nojiri and S.~D.~Odintsov, Unified cosmic history in modified gravity: from $F(R)$ theory to Lorentz non-invariant models, Phys. Rept. \textbf{505}, 59-144 (2011)

\bibitem{Carroll:2003wy} S.~M.~Carroll, V.~Duvvuri, M.~Trodden and M.~S.~Turner, Is cosmic speed-up due to new gravitational physics? Phys. Rev. D \textbf{70}, 043528 (2004)

\bibitem{Hu:2007nk} W.~Hu and I.~Sawicki, Models of $f(R)$ Cosmic Acceleration that Evade Solar-System Tests, Phys. Rev. D \textbf{76}, 064004 (2007)

\bibitem{Pradhan/2023} A. Pradhan, G. K. Goswami and A. Beesham, The reconstruction of constant jerk parameter with $f(R, T)$ gravity, J. High Energy Astrophys. {\bf 38}, 12 (2023)

\bibitem{Alvarenga:2013syu} F.~G.~Alvarenga, A.~de la Cruz-Dombriz, M.~J.~S.~Houndjo, M.~E.~Rodrigues and D.~S\'aez-G\'omez, Dynamics of scalar perturbations in $f(R,T)$ gravity, Phys. Rev. D \textbf{87}, 103526 (2013)

\bibitem{Sahoo:2017poz} P.~K.~Sahoo, P.~Sahoo and B.~K.~Bishi, Anisotropic cosmological models in $f(R,T)$ gravity with variable deceleration parameter, Int. J. Geom. Meth. Mod. Phys. \textbf{14}, 1750097 (2017)

\bibitem{Yousaf:2016lls} Z.~Yousaf, K.~Bamba and M.~Z.~u.~H.~Bhatti, Causes of Irregular Energy Density in $f(R,T)$ Gravity, Phys. Rev. D \textbf{93}, 124048 (2016)

\bibitem{Yousaf:2018jkb} Z.~Yousaf, M.~Z.~u.~H.~Bhatti and M.~Ilyas, Existence of compact structures in $f(R,T)$ gravity, Eur. Phys. J. C \textbf{78}, 307 (2018)

\bibitem{Yousaf:2016awj} Z.~Yousaf, K.~Bamba and M.~Z.~u.~H.~Bhatti, Influence of Modification of Gravity on the Dynamics of Radiating Spherical Fluids, Phys. Rev. D \textbf{93}, 064059 (2016)

\bibitem{Yousaf:2019zcb} Z.~Yousaf, K.~Bamba, M.~Z.~Bhatti and U.~Ghafoor, Charged Gravastars in Modified Gravity, Phys. Rev. D \textbf{100}, 024062 (2019)

\bibitem{Das:2016mxq} A.~Das, F.~Rahaman, B.~K.~Guha and S.~Ray, Compact stars in $f(R,\mathcal {T})$ gravity, Eur. Phys. J. C \textbf{76}, 654 (2016)

\bibitem{Yadav:2019zrw} A.~K.~Yadav, P.~K.~Sahoo and V.~Bhardwaj, Bulk viscous Bianchi-I embedded cosmological model in $f(R,T) = f1(R) + f2(R)f3(T)$ gravity, Mod. Phys. Lett. A \textbf{34}, 1950145 (2019)

\bibitem{Collins/1974} C. B. Collins, Tilting at cosmological singularities, Comm. Math. Phys. \textbf{39}, 131 (1974)

\bibitem{Coles/1994} P. Coles, G.F.R. Ellis, The case for an open universe, Nature \textbf{370}, 609 (1994)

\bibitem{Eriksen/2004} H. K. Eriksen, F. K. Hansen, A. J. Banday, K. M. Gorski, P. B. Lilje, Asymmetries in the CMB anisotropy field, Astrophys. J. \textbf{605}, 1420 (2004)

\bibitem{Yadav/2011}  A. K. Yadav, Some anisotropic dark energy models in Bianchi type-V space-time, Astrophys. Space Sci. \textbf{335}, 565 (2011)

\bibitem{Kumar/2011} S. Kumar, A. K. Yadav, Some Bianchi Type-V models of accelerating universe with dark energy, Mod. Phys. Lett. A \textbf{26}, 647 (2011)

\bibitem{Yadav/2012} A. K. Yadav, Bianchi—V string cosmological model and late time acceleration, Res. Astron. Astrophys. \textbf{12}, 1467 (2012)

\bibitem{Singh/2008} C. P. Singh, S. Ram, M. Zeyauddin, Bianchi type-V perfect fluid space-time models in general relativity, Astrophys. Space Sci. {\bf 315}, 181 (2008)

\bibitem{Goswami/2021} G. K. Goswami, A. K. Yadav, B. Mishra and S. K. Tripathy, Modeling of accelerating universe with bulk viscous fluid in Bianchi V space-time, Fortschr. Phys. {\bf 69}, 2100007 (2021)

\bibitem{Yadav/2014} A. K. Yadav, Bianchi-V string cosmology with power law expansion in $f(R, T)$ gravity Eur. Phys. J. Plus {\bf 129}, 194 (2014) 

\bibitem{Quzi/2023} S. Qazi et al., Classification of exact Bianchi type V cosmological solutions via conformal vector fields admitting energy conditions in  gravity, Results in Physics, {\bf 52}, 106710 (2023)

\bibitem{Gusu/2021} D. M. Gusu and M. V. Santhi, Analysis of Bianchi Type V Holographic Dark Energy Models in General Relativity and Lyra’s Geometry, Adv. High Energy Phys. {\bf 2021} 8818590 (2021)  

\bibitem{Barrow/2000} J. D. Barrow, Maartens and C. G. Tsagas, Cosmic acceleration in anisotropic universes, Phys. Rev. D {\bf 62}, 103502 (2000). 

\bibitem{Hewitt/2003} C. G. Hewitt, J. T. Horwood and J. Wainwright, Asymptotic Properties of the Bianchi V Cosmological Models, Class. Quant. Grav. {\bf 20}, 1743 (2003)





\bibitem{Lohakare/2022} S. V. Lohakare, B. Mishra, S. K. Maurya, Ksh. N. Singh, Analyzing the Geometrical and Dynamical Parameters of Modified Teleparallel- Gauss-Bonnet Model, arXiv: 2209.13197.

\bibitem{Yadav/2024JHEAP} A. K. Yadav et al., Reconstructing $f(Q)$ gravity from parameterization of the Hubble parameter and observational constraints, J. High Energy Astrophys. {\bf 43}, 114 (2024)

\bibitem{eBOSS:2020yzd} S. Alam et al. (eBOSS), Completed SDSS-IV extended Baryon Oscillation Spectroscopic Survey: Cosmological implications from two decades of spectroscopic surveys at the Apache Point Observatory, \href{https://doi.org/10.1103/PhysRevD.103.083533}{Phys. Rev. D 103, 083533 (2021)}.

\bibitem{Scolnic/2018} D. M. Scolnic et al., The Complete Light-curve Sample of Spectroscopically Confirmed SNe Ia from Pan-STARRS1 and Cosmological Constraints from the Combined Pantheon Sample, Astrophys. J. 859, 101 (2018).

\bibitem{Shahalam2015} M. Shahalam, S. Sami and A. Agarwa, Mon. Not. R. Astron. Soc. \textbf{448}, 2948 (2015).

\end{thebibliography}
\end{document}